# Focusing light with a deep parabolic mirror

N. Lindlein, M. Sondermann, R. Maiwald, H. Konermann, U. Peschel, G. Leuchs

*Institute of Optics, Information and Photonics (Max Planck Research Group),*
*Friedrich Alexander University of Erlangen-Nürnberg,*
*Staudtstr. 7/B2, 91058 Erlangen, Germany*
norbert.lindlein@optik.uni-erlangen.de

**Abstract:** The smallest possible focus is achieved when the focused wave front is the time reversed copy of the light wave packet emitted from a point in space (S. Quabis et al., Opt. Commun. 179 (2000) 1-7). The best physical implementation of such a pointlike sub-wavelength emitter is a single atom performing an electric dipole transition. In a former paper (N. Lindlein et al., Laser Phys. **17** (2007) 927-934) we showed how such a dipole-like radiant intensity distribution can be produced with the help of a deep parabolic mirror and appropriate shaping of the intensity of the radially polarized incident plane wave. Such a dipole wave only mimics the far field of a linear dipole and not the near field components. Therefore, in this paper, the electric energy density in the focus of a parabolic mirror is calculated using the Debye integral method. Additionally, a comparison with "conventional nearly $4\pi$" illumination using two high numerical aperture objectives is performed. The influence of aberrations due to a misalignment of the incident plane wave is discussed.

## References and links

1. P. Debye, „Das Verhalten von Lichtwellen in der Nähe eines Brennpunktes oder einer Brennlinie," Annu. Phys. **30**, 755-776 (1909).
2. B. Richards, E. Wolf, "Electromagnetic diffraction in optical systems II. Structure of the image field in an aplanatic system," Proc. R. Soc. A **253**, 358-379 (1959).
3. M. Mansuripur, "Distribution of light at and near the focus of high-numerical-aperture objectives," J. Opt. Soc. Am. A **3**, 2086-2093 (1986).
4. M. Mansuripur, "Distribution of light at and near the focus of high-numerical-aperture objectives: erratum," J. Opt. Soc. Am. A **10**, 382-383 (1993).
5. S. Hell, E.H.K. Stelzer, "Properties of a 4Pi confocal fluorescence microscope," J. Opt. Soc. Am. A **9**, 2159-2166 (1992).
6. N. Lindlein, R. Maiwald, H. Konermann, M. Sondermann, U. Peschel, G. Leuchs, "A new $4\pi$-geometry optimized for focussing onto an atom with a dipole-like radiation pattern," Laser Phys. **17**, 927-934 (2007).
7. C.J.R. Sheppard, A. Choudhury, J. Gannaway, "Electromagnetic field near the focus of wide-angular lens and mirror systems," Microwave Opt. Acoust. **1**, 129-132 (1977).
8. J. Stratton, L. Chu, "Diffraction theory of electromagnetic waves," Phys. Rev. **56**, 99-107 (1939).
9. P. Varga, P. Török, "Focusing of electromagnetic waves by paraboloid mirrors. I. Theory," J. Opt. Soc. Am. A **17**, 2081-2089 (2000).
10. P. Varga, P. Török, "Focusing of electromagnetic waves by paraboloid mirrors. II. Numerical results," J. Opt. Soc. Am. A **17**, 2090-2095 (2000).
11. S. Quabis, R. Dorn, M. Eberler, O. Glöckl, G. Leuchs, "Focusing light to a tighter spot," Opt. Commun. **179**, 1-7 (2000).
12. S. Quabis, R. Dorn, M. Eberler, O. Glöckl, G. Leuchs, "The focus of light-theoretical calculation and experimental tomographic reconstruction," Appl. Phys. B **B72**, 109-113 (2001).
13. R. Dorn, S. Quabis, G. Leuchs, "The focus of light - linear polarization breaks the rotational symmetry of the focal spot," J. Mod. Opt. **50**, 1917-1926 (2003).
14. R. Dorn, S. Quabis, G. Leuchs, "Sharper Focus for a Radially Polarized Light Beam," Phys. Rev. Lett. **91**, 233901 (2003).

15. M.A. Lieb, A.J. Meixner, "A high numerical aperture parabolic mirror as imaging device for confocal microscopy," Opt. Express **8**, 458-474 (2001).
16. N. Davidson, N. Bokor, "High-numerical-aperture focusing of radially polarized doughnut beams with a parabolic mirror and a flat diffractive lens," Opt. Lett. **29**, 1318-1320 (2004).
17. I.M. Bassett, "Limit to concentration by focusing," Optica Acta **33**, 279-286 (1986).
18. C.J.R. Shepppard, P. Török, "Electromagnetic field in the focal region of an electric dipole wave," Optik **104,** 175-177 (1997).
19. M. Sondermann, R. Maiwald, H. Konermann, N. Lindlein, U. Peschel, G. Leuchs, "Design of a mode converter for efficient light-atom coupling in free space," Appl. Phys. B **89,** 489-492 (2007).
20. D. McGloin, K. Dholakia, "Bessel beams: diffraction in a new light," Contemp. Phys. **46**, 15-28 (2005).

## 1. Introduction

The generation of the smallest possible focus is the aim of many scientific and industrial projects. It is well-known that for a given wavelength $\lambda$ and an optical system without aberrations the electric energy density in the focus depends on the aperture angle, the energy distribution of the radiant intensity propagating to the focus and the polarization. Following the standard way we calculate the electric energy density in the focus of a macroscopic optical system with a Fresnel number much larger than 1 by evaluating the Debye integral [1]. A vectorial formulation of this method for an aplanatic lens is given by Richards and Wolf [2] and similarly by Mansuripur [3,4]. The basic idea of this method is the following: Each geometrical optical ray which travels to the focus of the optical system represents a local plane wave and all these plane waves have to be superimposed coherently by taking into account the local polarization, the local amplitude and the local phase of each plane wave. To achieve not only a small focus in the lateral direction but also along the optical axis the light has to come from a nearly $4\pi$ solid angle. This results in a kind of standing wave pattern in the whole focal region. In confocal $4\pi$ microscopy such a nearly $4\pi$ illumination is produced by using two high numerical aperture aplanatic lenses arranged to produce overlapping foci but from opposite sides [5]. Additionally, the phase difference of both waves has to ensure positive interference at the focal point.

However, the use of two aplanatic lenses from opposite sides is not the only possibility to produce a nearly $4\pi$ illumination. The Cartesian ovoids like parabolic mirror, elliptical mirror or hyperbolic mirror are well-known for producing ideal imaging of one focal point onto the other one with up to $4\pi$ solid angle. Here, we will concentrate on the parabolic mirror which from a geometrical point of view produces an ideal focus for an incident plane wave front propagating exactly parallel to the optical axis. If the extension of the parabolic mirror tends to infinity the solid angle of the radiant intensity in the focal point reaches $4\pi$. The connection between the height $r$ of the incident ray and the aperture angle $\vartheta$ (i.e. angle between the optical axis and the light ray after reflection at the parabolic mirror) for a parabolic mirror with focal length $f$ is [6]:

$$r = 2f \tan\left(\frac{\vartheta}{2}\right) \qquad (1)$$

For example a parabolic mirror with a radius of $r=2f$ already reaches an aperture angle of $\vartheta=90°$, i.e. its numerical aperture is 1.0 in air. For parabolic mirrors with higher radius the aperture angle even exceeds 90° and may tend to 180°.

The calculation of the electric energy density in the focus of a parabolic mirror with a high numerical aperture close to 1.0 (in air, i.e. $\sin\vartheta=1.0$) and an incident linearly polarized plane wave using the Debye integral method as given by Richards and Wolf in [2] was demonstrated by Sheppard in [7]. There, it is also shown that this method is equivalent to applying the Stratton-Chu integral [8]. However, Varga and Török stated in [9,10] that the method of Richards and Wolf can only be used for aperture angles of $\vartheta\leq 90°$. For higher aperture angles they predicted different results, especially for an infinite parabolic mirror, by using the Stratton-Chu integral. However, we found a problem with the boundary conditions assumed for $\vartheta>90°$ in [9]. In equation (38) of this paper the authors assumed a plane wave with constant amplitude for the incident wave as well as for the emerging wave which is produced by a double reflection at the parabolic mirror. With simple geometrical optical calculations (also Varga and Török derived these boundary conditions assuming geometrical optics) it is however clear, that indeed the wave front (i.e. phase) of the doubly reflected wave is plane, but the amplitude is no longer constant. For an infinite aperture there would be a singularity on-axis of the amplitude because all the light power coming from zones of the incident wave with very large radii would be traveling nearly along the optical axis after two reflections at the parabolic mirror. So, the boundary conditions of Varga and Török for $\vartheta>90°$ are not appropriate and therefore also the results for very high aperture angles $\vartheta>90°$ should be reconsidered. In our opinion there is no reason why the method of Richards and Wolf should not work for a parabolic mirror with $\vartheta>90°$ as long as the aperture is not going to infinity. The method superposes plane waves along the direction of the geometrical optical rays traveling from the focal sphere to the focus, i.e. from a sphere around the focus whose radius of curvature is identical to the focal length. Of course, there is a limit in the case that the doubly reflected rays pass the focal point in a distance of just a few wavelengths, so that they cannot be neglected in the calculation. But, if this distance is much larger than several wavelengths there should not be any problem with the method of Richards and Wolf. With simple mathematics it can be shown that the height $r$ of the incident ray and the height $r'$ of the doubly reflected ray obey the equation $rr'=4f^2$. For parameters as in our experiment of focal length $f$=2.1 mm, wavelength $\lambda$=252 nm and ray height $r$≤100 mm (in the experiment we have only $r$≤10 mm, but $r$≤100 mm is used in one of our calculations) the closest distance of the doubly reflected rays to the optical axis will be $r'\geq 700\lambda$, which is far beyond the size of the focus with about 1 $\lambda$.

For a linearly polarized incident wave the focus will not be rotationally symmetric for high aperture angles as was shown in [7]. But, for a radially polarized incident wave the rotational invariance of the focus follows from symmetry arguments for both an aplanatic lens [11-14] and a parabolic mirror [15,16]. We will show in the following that for an aperture angle tending to 180° and a dipole-like radiant intensity [6] the ratio of the amplitude of the electric field in the focal point to the incident light power will reach a maximum compared to other configurations. This was generally shown by Bassett [17] using a multipole expansion of the fields and in analytic form by Sheppard and Török [18] using the diffraction integrals of Richards and Wolf. However, our numerical method does not explicitly assume an ideal dipole wave but a wave generated by the parabolic mirror by reflection of a quite arbitrary incident wave (with the only assumption that the incident wave front has to be nearly plane). In this way finite lateral extensions of the mirror, obscurations by a central hole in the mirror and misalignments of the incident wave, i.e. aberrations, are directly included.

The calculation of the electric energy density in the focus of a parabolic mirror or an aplanatic lens using our numerical implementation of the method of Richards and Wolf [2] is explained in section 2. In section 3 some numerical results for the parabolic mirror and a conventional 4π system are compared. Applications of the results, e.g. excitation of a single atom by a single photon, will be shortly discussed. Also, aberrations due to a misalignment of the incident plane wave will be taken into account.

## 2. Calculation of the electric (or magnetic) energy density in the focus of a parabolic mirror or an aplanatic lens

Following Richards and Wolf [2] a generalized version of the Debye integral is solved numerically. Plane waves along the geometrical optical rays are superimposed in the focal region taking into account the local polarization vector and possible phase shifts. However, different from the common simulation of an ideal lens it is now possible that a ray can have (small) aberrations so that it does not exactly hit the focus in the geometrical optical approximation. As long as the aberrations are small the ray is still assumed to represent a local plane wave which interferes with the other plane waves to obtain the electric (or magnetic) field in the focus.

*2.1 Ray tracing at a parabolic mirror*

The rays are calculated by ray tracing inside the parabolic mirror with focal length $f$ (see Fig. 1). The axis of the parabolic mirror is parallel to the z-axis, the vertex is at $z=f$ and the focus at $z=0$. So, a point with height $r$ and axial coordinate $z$ on the parabolic mirror surface is described by:

$$z(r) = -\frac{r^2}{4f} + f \quad \text{with } r = \sqrt{x^2 + y^2} \tag{2}$$

Using an implicit representation of the surface, the local surface normal unit vector $\mathbf{N}$ can be calculated:

$$G(\mathbf{r}) = z + \frac{x^2 + y^2}{4f} - f = 0 \Rightarrow \mathbf{N} = \frac{\nabla G}{|\nabla G|} = \frac{1}{\sqrt{1 + \frac{x^2 + y^2}{4f^2}}} \begin{pmatrix} \frac{x}{2f} \\ \frac{y}{2f} \\ 1 \end{pmatrix} \tag{3}$$

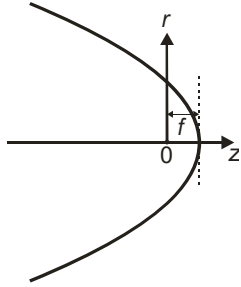

Fig. 1. Parameters of the parabolic mirror with focal length $f$. The origin of the coordinate system is at the point 0 which is also the focus of the mirror.

The incident ray is described by the starting point $\mathbf{p}$, unit direction vector $\mathbf{e}$, optical path length difference OPD at the starting point and vacuum wavelength $\lambda$ or frequency $\nu=c/\lambda$. Additionally, it has a polarization vector $\mathbf{P}$ which is complex valued and which is connected to the complex (time-independent) electric field vector $\mathbf{E}$ of the assigned "local plane wave" by the equation:

$$\mathbf{E}(\mathbf{r}) = \frac{1}{dF}\mathbf{P}\exp\left(i\frac{2\pi n}{\lambda}\mathbf{e}\cdot(\mathbf{r}-\mathbf{p}) + i\frac{2\pi}{\lambda}\text{OPD}\right) = \frac{1}{dF}\mathbf{P}\exp\left(i\frac{2\pi\nu}{c}[n\mathbf{e}\cdot(\mathbf{r}-\mathbf{p}) + \text{OPD}]\right) \tag{4}$$

Here, d$F$ is the surface element which is associated with each ray of the incident plane wave and thus with the numerical sampling density (see Fig. 2). This means especially, that the modulus of the polarization vectors of the incident plane waves associated with each ray has to be calculated according to:

$$P = E\,dF \qquad (5)$$

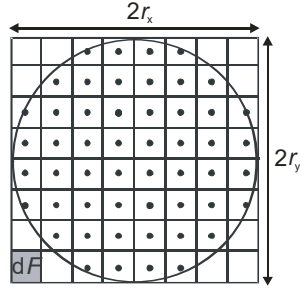

Fig. 2. Sampling of the incident wave by rays (dots representing the points of intersection of the rays with the entrance pupil). In the figure, the number of rays along each axis is $N_{ray}=8$.

For an aperture with radius $r_x$ in x-direction and radius $r_y$ in y-direction which is sampled by $N_{ray}$ rays along each lateral axis (i.e. $N_{ray}^2$ in total since an equidistant sampling on a rectangular base surface is assumed) the surface element $dF$ is (see Fig. 2):

$$dF = \frac{4\,r_x r_y}{N_{ray}^2} \qquad (6)$$

So, the point of intersection of the incident ray, which represents a local plane wave, with the paraboloid is:

$$G(\mathbf{p}+\mu\mathbf{e}) = 0 \;\Rightarrow\; \left[\frac{e_x^2+e_y^2}{4f}\right]\mu^2 + \left[e_z + \frac{p_x e_x + p_y e_y}{2f}\right]\mu + \left[p_z + \frac{p_x^2+p_y^2}{4f} - f\right] = 0 \qquad (7)$$

Here, $\mu$ is the path length from the starting point of the ray to the point of intersection with the parabolic mirror. With the path length $\mu$ calculated, the point of intersection with the surface, which is then the new starting point of the reflected ray, is:

$$\mathbf{p}' = \mathbf{p} + \mu\mathbf{e} \qquad (8)$$

The direction vector $\mathbf{e}'$ of the reflected ray is given by the reflection law using the surface normal vector $\mathbf{N}$ of equation (3):

$$\mathbf{e}' = \mathbf{e} - 2(\mathbf{e}\cdot\mathbf{N})\mathbf{N} \qquad (9)$$

With this the new direction vector of the reflected ray is known and the optical path length at the new starting point is:

$$\mathrm{OPD}' = \mathrm{OPD} + n\mu \qquad (10)$$

Here, $n$ is the refractive index of the medium which is in our case air, i.e. $n=1$.

The direction of the polarization vector $\mathbf{P}$ changes upon reflection as well. For an ideal conducting mirror we can assume that the component of the reflected electric field parallel to the surface has the inverse value of the incident component. Then, the total electric field parallel to the surface is zero and no electric currents are induced. The component perpendicular to the surface is not changed. This results in:

$$\mathbf{P}' = (\mathbf{P}\cdot\mathbf{N})\mathbf{N} - [\mathbf{P} - (\mathbf{P}\cdot\mathbf{N})\mathbf{N}] = -\mathbf{P} + 2(\mathbf{P}\cdot\mathbf{N})\mathbf{N} \qquad (11)$$

*2.2 Superposition of the local plane waves in the focus*

The modulus of the electric field vector and therefore of the polarization vector has to be calculated for all local plane waves assigned to the rays on the focal sphere. Only if the electric field vectors of the local plane waves are known in a constant distance from the focus which is much larger than the extension of the focus, the electric field vectors can be added in the focus correctly. Due to energy conservation there is an apodisation factor $g$, depending on the focussing optical element/system, which links the modulus of the polarization vector of the incident ray $P_{in}$ to the modulus of the polarization vector of the ray on the focal sphere $P_{out}$. This factor is calculated in subsection 2.3.

In the focal region the complete electric energy density can be calculated using a coherent superposition of the different plane waves which are assigned to the rays numbered by the index $j$. Each plane wave is of the form (4), so that we have:

$$\mathbf{E}_{total}(\mathbf{r}) = \frac{1}{i\lambda f} \int_{focal\,sphere} \mathbf{E}\,dF = \frac{1}{i\lambda f} \sum_j g_j \mathbf{P'}_j \exp\left(i\frac{2\pi\nu}{c}\left[n\mathbf{e'}_j \cdot (\mathbf{r} - \mathbf{p'}_j) + \mathrm{OPD'}_j\right]\right) \quad (12)$$

Here, the integral is taken over the surface of the focal sphere with radius $f$. Numerically, this is replaced by a sum over local plane waves which are assigned to the rays. The factor $g_j$ is the apodisation factor of ray number $j$. The replacement of $|\mathbf{E}\,dF|$ by $|g\mathbf{P'}|$ (sampled on the focal sphere with surface element $dF$) is due to the fact that, as mentioned before, the modulus $E$ of the electric vector is in our numerical simulation method the density of the modulus of the polarization vector:

$$E = \frac{P_{focalsphere}}{dF} = \frac{gP'}{dF} \quad (13)$$

Here, $\mathbf{P'}$ takes only into account the change of the polarization vector directly at the optical element (taking for example equation (11) for reflection at an ideal mirror), so that the change of the modulus of the polarization vector, which has to be calculated at the position of the focal sphere, is taken into account by the apodisation factor $g$.

This method should work as long as the aberrations are small. The modulus of the electric vector in the focus can be calculated relative to the modulus of the electric vector of the incident wave which is normalized in such a way that the maximum value of the electric vector is one unit. However, the electric vector can also be calculated in absolute physical units relative to the incident light power.

A precondition for our method is that the focal length $f$ is much larger than the wavelength $\lambda$ and that only points in the vicinity of the focus are regarded.

*2.3 Calculation of the apodisation factor for the parabolic mirror*

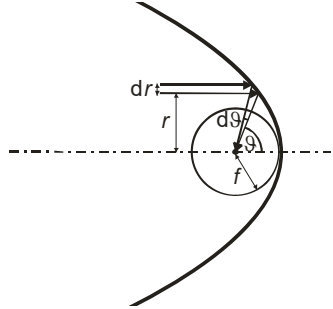

Fig. 3. Scheme explaining the relation between an incident ray with height $r$ and a deflected ray on the focal sphere with angle $\vartheta$.

The apodisation factor $g$ is calculated by using energy conservation between the incident and deflected light (assuming no absorption). So, the light power which is in an annulus with radius $r$ and thickness $dr$ of the incident plane wave has to be the same as in an annulus on the focal sphere with angle $\vartheta$ with the optical axis, radius $f$ (focal length) and "angular thickness" $d\vartheta$ (see Fig. 3). The surface element $dF_{in}$ of the annulus on the incident plane wave is:

$$dF_{in} = 2\pi r\, dr \tag{14}$$

The surface element d$F_{out}$ on the focal sphere is:

$$dF_{out} = 2\pi f^2 \sin\vartheta\, d\vartheta \tag{15}$$

Due to energy conservation the relation between the intensities $I_{in}$ and $I_{out}$ of the local plane waves (which propagate both perpendicular to the surface elements so that the concept of intensity is appropriate here) is:

$$I_{in}dF_{in} = I_{out}dF_{out} \Rightarrow \frac{E_{out}}{E_{in}} = \frac{\sqrt{I_{out}}}{\sqrt{I_{in}}} = \sqrt{\frac{dF_{in}}{dF_{out}}} \tag{16}$$

Here, $E_{in}$ and $E_{out}$ are the moduli of the electric vectors of the incident and deflected local plane waves.

As mentioned before, in a numerical simulation where non-equidistant rays/local plane waves are added, the electric vector **E** is not directly proportional to the polarization vector **P**, but the electric vector is proportional to the polarization vector per surface element:

$$E_{in} = \frac{P_{in}}{dF_{in}};\quad E_{out} = \frac{P_{out}}{dF_{out}};\quad \Rightarrow g = \frac{P_{out}}{P_{in}} = \frac{E_{out}dF_{out}}{E_{in}dF_{in}} = \sqrt{\frac{dF_{out}}{dF_{in}}} \tag{17}$$

So, after reflection at the parabolic mirror using equation (11) the polarization vector has to be multiplied with the apodisation factor $g$. This factor can be calculated using geometrical optical methods. From equation (1) we know the relation between the height $r$ of the incident ray which propagates parallel to the optical axis and the reflected ray with angle $\vartheta$ relative to the optical axis.

Using equations (1), (14), (15) and (17) the apodisation factor for the parabolic mirror is:

$$r = 2f\tan\left(\frac{\vartheta}{2}\right) \Rightarrow dr = f\frac{1}{\cos^2\left(\frac{\vartheta}{2}\right)}d\vartheta \Rightarrow$$

$$g = \sqrt{\frac{dF_{out}}{dF_{in}}} = \sqrt{\frac{f^2\sin\vartheta\, d\vartheta}{r\, dr}} = \sqrt{\frac{2\sin\left(\frac{\vartheta}{2}\right)\cos\left(\frac{\vartheta}{2}\right)}{2\sin\left(\frac{\vartheta}{2}\right)\cos^{-3}\left(\frac{\vartheta}{2}\right)}} = \cos^2\left(\frac{\vartheta}{2}\right) = \frac{1}{2}(1+\cos\vartheta) \tag{18}$$

Note that in our numerical simulation the apodisation factor is defined as the ratio $P_{out}/P_{in}$ (equation (17)), whereas in the literature it is mostly defined as $E_{out}/E_{in}$ (see equation (16)). This explains why our apodisation factor is the reciprocal value of that given in the literature (see for example [7]).

Our considerations assume an ideally reflecting mirror without losses. However, it is straightforward to include losses of a real metallic mirror by taking into account the Fresnel equations for metallic surfaces. Then equation (11) has to be replaced by an expression that corrects the amplitude and phase of the polarization vector **P** accordingly.

*2.4 Summary of calculating the electric energy density in the focus of a parabolic mirror*

In summary the Debye integral can be evaluated by using equation (12) where equations (7)-(11) and (18) are used to calculate the starting point $\mathbf{p'}_j$, direction vector $\mathbf{e'}_j$, optical path length difference at the starting point $OPD'_j$, polarization vector $\mathbf{P'}_j$ and apodisation factor $g_j$ of the reflected ray/plane wave which propagates from the point of intersection with the parabolic mirror to the focus. Notice that $\mathbf{p'}_j$, $\mathbf{e'}_j$, $OPD'_j$ and $\mathbf{P'}_j$ are directly calculated at the point of intersection of the incident ray with the parabolic mirror, but $g_j$ takes into account the relative strength of the amplitude of the respective local plane wave on the focal sphere. If the incident wave is not exactly parallel to the optical axis or if it has a small curvature, the reflected rays will not exactly point to the focus. However, our equations are still valid as long as the aberrations are small. Surface deviations $\Delta S$ of the parabolic mirror can also be included by multiplying the polarization vector after reflection with a phase factor $\exp(4\pi i\Delta S/\lambda)$. Here, $\Delta S$ are the surface deviations along the surface normal and in the phase factor it is taken into account that these surface deviations are passed twice.

*2.5 Calculation of the electric energy density in the focus of an aplanatic lens*

In the case of an aplanatic lens instead of the parabolic mirror, equation (12) can also be used, but with different apodisation factor $g$ and ray parameters. The analogous model of an ideal aplanatic lens, which replaces a quite complex sequence of many lenses by one "virtual optical element", is that all incident rays seem to be deflected directly at the focal sphere in such a way that they propagate to the focus with identical optical path lengths. This model is identical to the sine condition $r = f\sin\vartheta$, which connects the ray height $r$ of the incident ray with the ray angle $\vartheta$ of the ray propagating to the focus. For an ideal aplanatic lens the focus can be taken as the reference point where the deflected rays start. So, it is $\mathbf{p}'=0$ and $\text{OPD}'=0$. The direction vector $\mathbf{e}'$ points from the focal sphere to the focus. The polarization vector $\mathbf{P}'$ is calculated from the polarization vector $\mathbf{P}$ of the incident ray in such a way that the component perpendicular to the plane of deflection, which is spanned by the direction vector $\mathbf{e}$ of the incident ray and the direction vector $\mathbf{e}'$ of the deflected ray, remains unchanged, whereas the component of $\mathbf{P}$ in the deflection plane has to be rotated in this plane so that it is perpendicular to $\mathbf{e}'$. The following unit vectors are introduced: $\mathbf{e}_\perp$ is perpendicular to the plane of deflection, $\mathbf{e}_\parallel$ is in the plane of deflection perpendicular to $\mathbf{e}$ and $\mathbf{e}'_\parallel$ is in the plane of deflection perpendicular to $\mathbf{e}'$. Then, we can summarize for the polarization vector $\mathbf{P}'$ of the deflected ray:

$$\mathbf{e}_\perp = \frac{\mathbf{e}\times\mathbf{e}'}{|\mathbf{e}\times\mathbf{e}'|}; \quad \mathbf{e}_\parallel = \mathbf{e}_\perp \times \mathbf{e} \quad \text{and} \quad \mathbf{e}'_\parallel = \mathbf{e}_\perp \times \mathbf{e}' \qquad (19)$$
$$\Rightarrow \mathbf{P}' = \mathbf{P} - (\mathbf{P}\cdot\mathbf{e}_\parallel)\mathbf{e}_\parallel + (\mathbf{P}\cdot\mathbf{e}_\parallel)\mathbf{e}'_\parallel$$

Equation (19) is not valid for a ray collinear with the optical axis. But, this ray is not deflected at all and so the polarization vector is unchanged.

The apodisation factor $g$ of the aplanatic lens is calculated analogous to equation (18) by taking the relation between the height $r$ of the incident ray and the angle $\vartheta$ of the deflected ray of an aplanatic lens (sine condition for object point at infinity):

$$r = f\sin\vartheta \Rightarrow dr = f\cos\vartheta\, d\vartheta \Rightarrow g = \sqrt{\frac{dF_{out}}{dF_{in}}} = \sqrt{\frac{f^2\sin\vartheta\, d\vartheta}{r\, dr}} = \frac{1}{\sqrt{\cos\vartheta}} \qquad (20)$$

This allows to calculate the electric energy density near the focus of an ideal aplanatic lens. A conventional 4π system with two aplanatic lenses can also be simulated by coherently adding the resulting electric vectors of both lenses in the focal region.

*2.6 Calculation of the magnetic energy density in the focus*

The magnetic vector $\mathbf{H}_{total}$ in the focus can be calculated with the same method as for the electric energy density if the electric vectors in the equations are replaced by magnetic vectors. Then, the polarization vectors represent the magnetic field $\mathbf{H}$ which has to be perpendicular to the electric field $\mathbf{E}$ and the direction of propagation $\mathbf{e}$ for a plane wave, i.e. $\mathbf{H} \propto \mathbf{e}\times\mathbf{E}$. Only, for reflection at the parabolic mirror we have to use another equation as (11) since upon reflection the normal component of the magnetic vector changes sign whereas the tangential component remains constant:

$$\mathbf{P}' = -(\mathbf{P}\cdot\mathbf{N})\mathbf{N} + [\mathbf{P} - (\mathbf{P}\cdot\mathbf{N})\mathbf{N}] = \mathbf{P} - 2(\mathbf{P}\cdot\mathbf{N})\mathbf{N} \qquad (21)$$

So, in the case of the parabolic mirror equation (11) has to be replaced by equation (21) if the magnetic field is treated instead of the electric one. For the ideal aplanatic lens equation (19) is also valid for the magnetic field.

**3. Numerical simulations of the parabolic mirror and the conventional 4π system**

The characteristic property of a dipole radiation is that the majority of the light is emitted in a plane perpendicular to the dipole axis and no light is emitted parallel to the dipole axis. The exact mathematical expression for the radiant intensity $R_{dipole}$ of a dipole radiation in the far field is:

$$R_{dipole}(\vartheta) = R_0 \sin^2 \vartheta \qquad (22)$$

Again, $\vartheta$ is the angle with the optical axis which coincides with the dipole axis. The ideal dipole radiation has its maximum at an angle of $\vartheta=90°$. Therefore, it is clear that a "conventional 4π-setup" with two high aperture objectives which illuminate the focus along the optical axis cannot produce such a dipole radiation. Even, if the numerical aperture (in air) is 0.95, what is about the maximum value in practice, the maximum aperture angle $\vartheta$ is just:

$$NA = \sin \vartheta = 0.95 \Rightarrow \vartheta = 71.8° \qquad (23)$$

This means that a large part of the dipole radiation is missing even if the radiant intensity in the other parts would be shaped according to a dipole pattern. This can be proven by calculating the total light power $W$ which is contained in a dipole radiation in a range between the minimum aperture angle $\vartheta_{min}$ and the maximum aperture angle $\vartheta_{max}$:

$$W = 2\pi \int_{\vartheta_{min}}^{\vartheta_{max}} R_{dipole}(\vartheta) \sin \vartheta \, d\vartheta = 2\pi R_0 \left[ \cos \vartheta_{min} - \cos \vartheta_{max} + \frac{1}{3} \cos^3 \vartheta_{max} - \frac{1}{3} \cos^3 \vartheta_{min} \right] \qquad (24)$$

In a half space with solid angle 2π, i.e. $\vartheta_{min}=0°$ and $\vartheta_{max}=90°$, there is $W_{2\pi}=4\pi R_0/3$ total light power. However, if $\vartheta_{min}=0°$ and $\vartheta_{max}=71.8°$ there is only 55% of the total light power of the 2π solid angle contained in the resulting cone. From that follows that it is necessary to have another optical device to produce in the focus the radiant intensity of a real dipole radiation. A parabolic mirror which is illuminated by a well-shaped incident wave is such a device which allows aperture angles of 90° and more.

In the following, some numerical calculations of the electric energy density in the focus of a parabolic mirror and of a conventional 4π system are compared. In the case of the parabolic mirror it is assumed that the intensity/irradiance $I$ of the plane wave front which is focused by the parabolic mirror is shaped in such a way that it results in the radiant intensity of a dipole [6] and only the minimum and maximum aperture angles are different from $\vartheta_{min}=0°$ and $\vartheta_{max}=180°$, respectively:

$$I(r) = I_0 \frac{(r/f)^2}{\left(1 + \frac{(r/f)^2}{4}\right)^4} \qquad (25)$$

Again, $f$ is the focal length of the parabolic mirror and $r$ is the lateral coordinate, i.e. the distance from the optical axis. The polarization of the incident wave is radial to obtain the dipole-like radiant intensity after reflection at the mirror.

For the conventional 4π system the intensity distribution of the incident plane wave front for radial polarization is a doughnut mode with a beam parameter $w$:

$$I(r) = I'_0 \, r^2 \exp\left(-2 \frac{r^2}{w^2}\right) \qquad (26)$$

Moreover, the beam parameter $w$ is chosen in such a way that the maximum value of the intensity of the doughnut mode is at the rim of the aperture at $r=r_{aperture}$.

The case of linear polarization of the incident wave is also simulated for the conventional 4π system. Then, the incident wave is a perfect plane wave with constant intensity over the aperture.

*3.1 Parabolic mirror*

For the numerical simulations of the parabolic mirror a wavelength of $\lambda$=252 nm and a focal length of $f$=2.1 mm is used. Three parabolic mirrors with different aperture radii were simulated. In the first case, there is a central hole in the parabolic mirror which is used to put a thin pin through it for fixing an object at the focus (for example a sphere for testing the parabolic mirror or part of an ion trap for fixing an ion in the focus). The two other parabolic mirrors have no central hole and increased aperture sizes. By using equation (1) the minimum and maximum aperture angles $\vartheta_{min}$ and $\vartheta_{max}$ were calculated for all three cases. With the help of equation (24) the ratio $\eta$ of the total light power which is collected by the respective parabolic mirror compared to the light power of an infinite full parabolic mirror using the dipole radiation is obtained. The parameters of the mirrors are listed in table 1.

**Table 1: Parameters of the simulated parabolic mirrors which all have a focal length of $f$=2.1 mm.**

|  | Mirror 1 | Mirror 2 | Mirror 3 |
|---|---|---|---|
| $r_{min}$ | 0.75 mm | 0 mm | 0 mm |
| $r_{aperture}$ | 10 mm | 20 mm | 100 mm |
| $\vartheta_{min}$ | 20.2° | 0° | 0° |
| $\vartheta_{max}$ | 134.4° | 156.3° | 175.2° |
| $\eta$ | 0.936 | 0.995 | 0.99999 |

Mirror 1 is the parabolic mirror which is tested experimentally in our laboratory. The two other mirrors are test cases for the simulations to cover a range of parameters. Mirror 3 can be assumed for our case as being an infinitely extended parabolic mirror because it collects 99.999% of the total light power which is contained in a dipole wave.

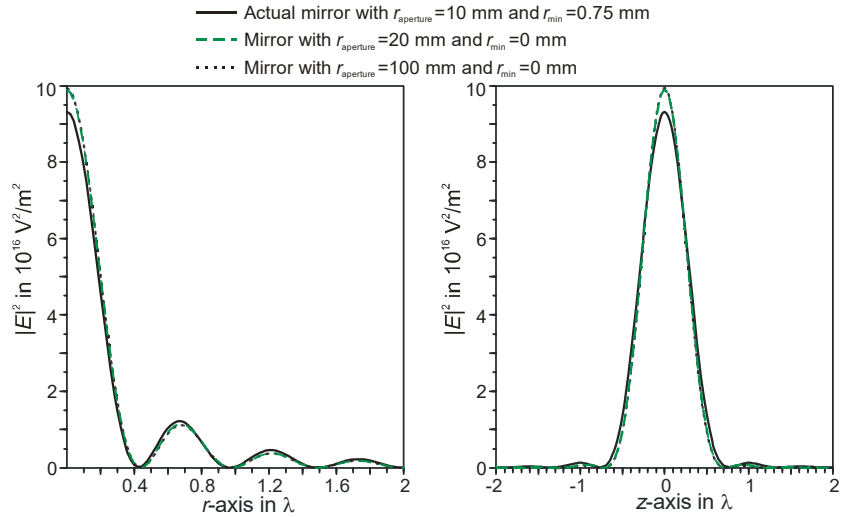

Fig. 4. Lateral section (left) and axial section (right) of the square of the electric field in the focal region of three parabolic mirrors with different aperture sizes assuming an incident light power of 1 W in each case. Note that the green dashed and black dotted curves nearly coincide showing that mirror 2 and 3 would produce in practice a nearly ideal dipole wave with 4π solid angle.

The electric energy density near the focus is evaluated on a field with 201x201 samples and a field radius of 2 wavelengths assuming an aberration-free parabolic mirror. This field can be either in the x-y-plane (focal plane) or in a plane containing the optical axis (i.e. x-z- or y-z-plane). In principle, we can calculate the electric energy density in a small volume around the focus. The number of rays/plane waves for sampling the wave front is 200x200 on an orthogonal grid. Since the aperture is annular (for mirror 1) or circular, the effective number of sampling points is less (about a factor $\left(r_{aperture}^2 - r_{min}^2\right)\pi / 4 r_{aperture}^2 = 0.78$ for mirror 1).

Figure 4 shows lateral and axial sections of the square of the total electric field for the three different mirrors assuming that in each case the incident light power is 1 W. It can be seen that there is a small difference between mirror 1 and the two other mirrors, but nearly no difference between mirrors 2 and 3. Also the difference between mirror 1, which will be used in our experiments, and the two other "virtual" mirrors is so small that in practice mirror 1 (assuming there are no surface deviations causing aberrations) will produce a nearly ideal focus of a dipole-like radiation converging to the focus.

If the different components of the electric field in the focal region are considered separately, there will be mainly a strong axial component of the electric field at the focal plane, whereas the lateral components are nearly vanishing (see Fig. 5 for the case of mirror 1), and there will be due to symmetry reasons a pure axial component along the optical axis. The ratios of the maximum values of the axial component to the lateral component of the electric energy density in the focal plane are 400, $10^4$ and $3 \cdot 10^6$ for the mirrors 1, 2 and 3, respectively. So, there will be a pure axial component of the electric field in the focal plane for an infinite ideal parabolic mirror illuminated with a plane wave front and the ideally shaped intensity distribution producing a dipole-like radiation after reflection. For our finite mirror 1 the lateral components of the electric field in the focal plane can likewise be neglected (see Fig. 5). Outside the focal plane there are also lateral components of the electric field for an infinite parabolic mirror which are nearly identical to that of Fig. 5.

It should also be noted that the maximum of the square of the electric field in the focus is according to Fig. 4 for mirror 3, i.e. the ideal dipole wave, and $W$=1 W incident light power: $|E|^2 = 9.94 \cdot 10^{16}$ V$^2$/m$^2$. This is the exact same value as calculated by Bassett [17] to be the maximum value which can be achieved by focusing at all:

$$|E|^2_{max} = \frac{16\pi}{3\varepsilon_0 c \lambda^2} W = 9.94 \cdot 10^{16} \text{ V}^2/\text{m}^2 \qquad (27)$$

With dielectric constant of vacuum $\varepsilon_0$, speed of light in vacuum $c$ and wavelength $\lambda$=252 nm.

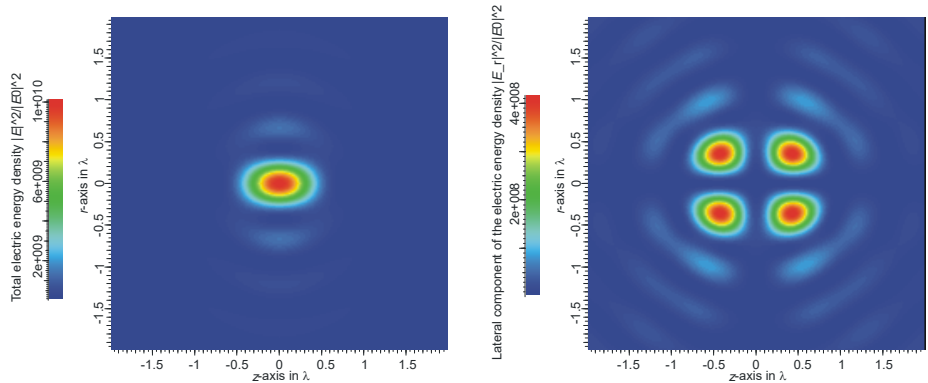

Fig. 5. Comparison of the total electric energy density (left) and the lateral component of the electric energy density only (right) in a plane containing the optical axis (z-axis horizontally) and a lateral axis (r-axis vertically). Simulated for the parabolic mirror number 1 using radial polarization and the ideal intensity distribution of the incident wave producing a dipole wave. The electric energy

density is given relative to the maximum electric energy density of the incident wave. The lateral scale is in multiples of the wavelength (λ=252 nm) of the used light. Note the different scales of the electric energy density in both figures.

*3.2 Conventional 4π setup*

Here, a conventional 4π setup consisting of two aplanatic objectives with a common focus is simulated. We assume that the two aplanatic objectives of the 4π setup have a numerical aperture of $\sin\vartheta=0.95$ (in air) which is about the maximum numerical aperture that can be achieved in practice. The focal length is again $f=2.1$ mm as in the case of the parabolic mirror. Of course, since the objectives fulfil the sine condition the maximum aperture radius is now only $f \sin\vartheta = 1.995$ mm, whereas it was 10 mm for the parabolic mirror 1. So, the maximum values of the electric energy density which are given in the following Fig. 7 and 8 relative to the maximum value of the incident wave will be smaller as for the parabolic mirror due to the smaller amount of collected light power and cannot be compared directly. Moreover, also the intensity distribution of the incident wave is different from that of the parabolic mirror. Therefore, later on in Fig. 9 we will again present the square of the electric field in physical units for a total incident light power of 1 W in all cases.

The simulations are made for four cases: (i) radial polarization and doughnut mode intensity of the incident waves (maximum of the intensity at the rim of the aperture) for a full circular aperture, (ii) the same incident waves but with an annular aperture where the inner radius of the aperture is at 90% of the maximum aperture radius, (iii) incident plane waves with linear polarization in x-direction and constant intensity. Case (iv) again uses radial polarization, but the intensity of the incident wave is shaped in such a way that behind the objectives a dipole-like wave is generated. This dipole-like wave has no contributions at aperture angles between 71.8° and 108.2° where a real dipole wave has the maximum of its radiant intensity. In all cases, the wave fronts of the incident waves are plane concerning the phase.

In order to have positive interference in the common focal plane of both objectives, the phase of the two plane wave fronts, incident onto the two objectives, must have a relative phase shift of π in the case of radial polarization. Without this phase shift the radially polarized electric field vectors would cancel each other in the focal plane (see Fig. 6a). But, with this phase shift of π there is a strong axial component of the electric field in the focal plane which increases with increasing numerical aperture (see Fig. 6b). For linear polarization of the incident waves both waves have to be exactly in phase in order to have positive interference in the focus.

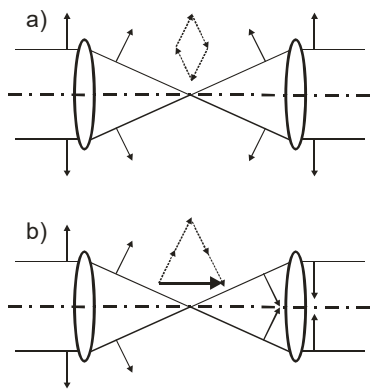

Fig. 6. Scheme of a conventional 4π setup illuminated with radial polarization showing the formation of the electric field in the focus. In a) there is no phase shift between the incident waves, whereas in b) there is a phase shift of π.

Figure 7 shows the total electric energy density and the lateral component of the electric energy density for the conventional 4π setup with full or annular aperture and radial polarization. By comparing Fig. 5 and 7 it can be seen that in all cases of radial polarization the lateral component of the electric field in the focal plane is nearly or exactly zero. This is clear due to symmetry reasons if the radiant intensity coming from the left half space is identical to that coming from the right half space. In the case of parabolic mirror 1 the lateral component of the electric field is only nearly zero since the incident radiant intensity is not completely symmetric in the two half spaces. For the conventional 4π setup it is completely symmetric and thus the lateral component is exactly zero in the focal plane.

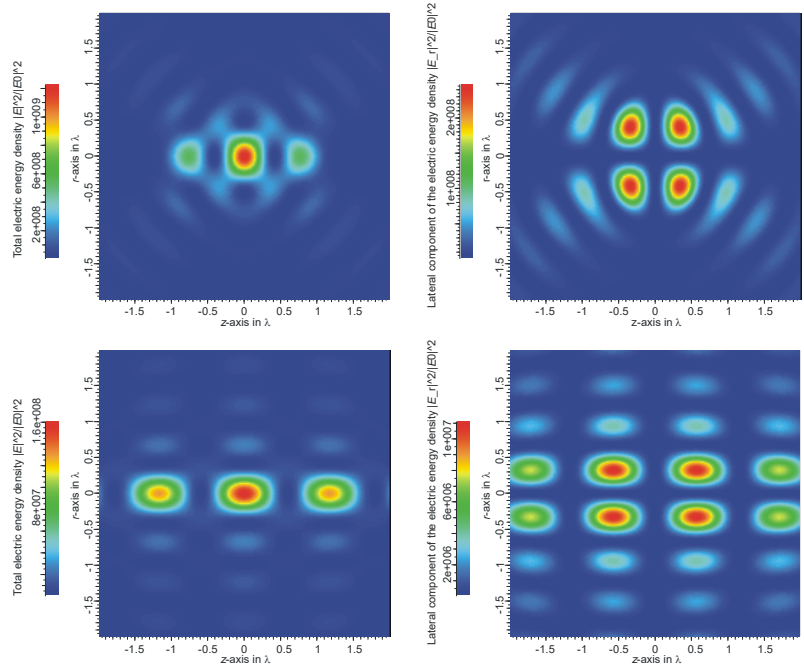

Fig. 7. Simulation for the conventional 4π setup with radial polarization and a full aperture (top figures) and an annular aperture with 90% of the radius blocked (bottom figures). The figure shows the comparison of the total electric energy density (left) and the lateral component of the electric energy density only (right) in a plane containing the optical axis (z-axis horizontally) and a lateral axis (r-axis vertically).. The lateral scale is in multiples of the wavelength (λ=252 nm) of the used light. Note the different scales for the electric energy density in the figures.

Figure 8 shows the total electric energy density (left figures) in the xz- and yz-plane for the case of linear polarization in x-direction and a full circular aperture (an annular aperture is not simulated here). In both cases the component of the electric field in x-direction (along the direction of polarization) is the dominant component and in the yz-plane it is also the only component (i.e. total electric energy density = electric energy density due to the x-component). Besides the x-component there is also a smaller axial component in the xz-plane contributing to the total electric energy density. The right figure of the top row shows also the electric energy density of the axial component alone.

So, contrary to radial polarization, where there is only an axial component of the electric field in the focal plane, there is only the lateral x-component of the electric field in the focal plane for linear polarization of the incident waves (polarization in x-direction). For linear polarization there is also a slightly smaller lateral extension of the focus in y-direction (for incident polarization in x-direction) than in x-direction.

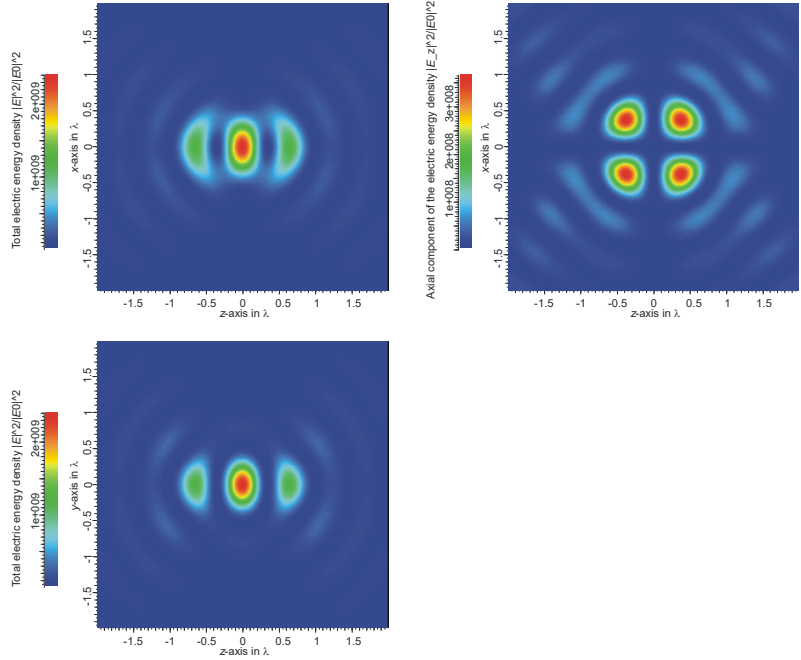

Fig. 8. Simulation for the conventional 4π setup with linear polarization in x-direction and a circular aperture. The figure shows the comparison of the total electric energy density (left) and the axial component of the electric energy density only (right) in a plane containing the optical axis (z-axis horizontally) and one of the lateral axes (x-axis top figures, y-axis bottom figure; vertical axes, respectively). The lateral scale is in multiples of the wavelength of the used light. Note the different scales of the electric energy density in the left and right top figures.

*3.3 Comparison between parabolic mirror with dipole-like radiant intensity and conventional 4π setup*

Figure 9 shows a direct comparison of the square of the electric field using a parabolic mirror or a conventional 4π setup with different types of polarization and different types of the intensity distribution of the incident wave. In all cases the incident light power is 1 W so that the maximum values of the curves can be compared directly.

The different curves can now be evaluated concerning different applications. Here, we want to focus on an experiment which is just built up at our institute to study the absorption of a single atom in free space if the incident light mimics the radiant intensity of a dipole wave which is the far field wave emitted by a linear dipole transition in an atom [19]. The idea is that the atom should absorb a single photon with the highest possible probability (i.e. nearly one) if the inverse spatio-temporal distribution of a spontaneously emitted photon is generated. Since the atom will only "see" the electric field of the incident wave at its position, the maximum value of the electric field in the focus relative to the incident light power should be as high as possible. Furthermore, only the field component parallel to the dipole moment of the atomic transition interacts with the atom. Hence, if the quantization axis is fixed along the optical axis of the mirror/aplanatic lenses (e.g. by applying external magnetic fields) and a transition involving linearly polarized light is driven, one has to maximize the field component parallel to the optical axis in the focus. Therefore, the height of the maximum of the curves in Fig. 9 directly expresses the appropriateness for our experiment, since it is determined solely by the relevant field component. It can clearly be seen that the parabolic mirror using radial polarization and an incident light intensity producing a dipole-like radiant intensity after reflection is the best setup. Also, if the incident light has a doughnut mode intensity, which is the normal case for radial polarization without any additional beam shaping, the electric field in the focus will be higher than values which can be achieved with conventional 4π setups. For the conventional 4π setups using two objectives the best case is the simple case of linear polarization and constant intensity of the incident wave (note that in this case the quantization axis of the atom for the absorption experiment has to be in x-direction, i.e. parallel to the incident polarization). Radial polarization produces lower maxima of the electric field.

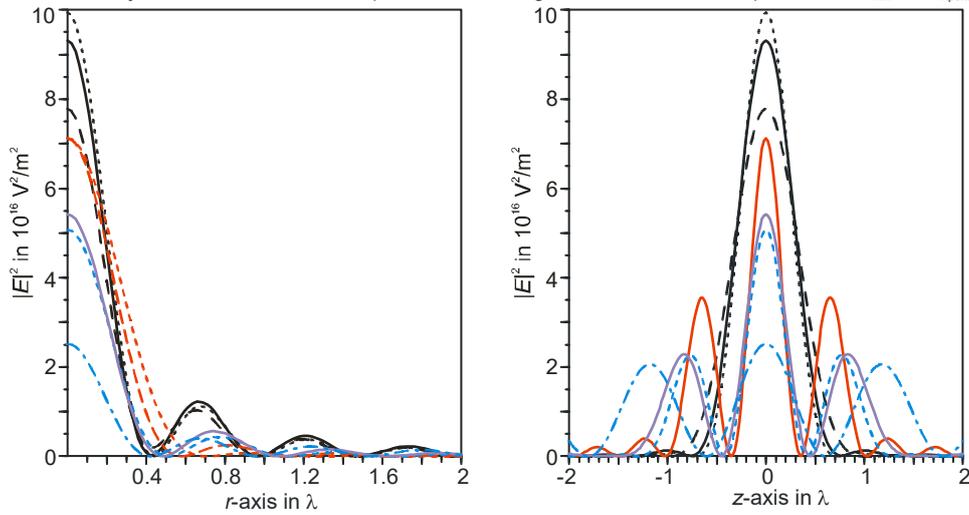

Fig. 9. Comparison of the square of the electric field in the focus for the different setups assuming in all cases an incident light power of 1 W (left: lateral section, right: axial section):
a) (nearly) infinite parabolic mirror with radial polarization and incident wave producing an ideal dipole-like radiant intensity (dotted black curve),
b) finite parabolic mirror as it is used in our experiment with radial polarization and incident wave producing an ideal dipole-like radiant intensity (solid black curve),
c) finite parabolic mirror as it is used in our experiment with radial polarization, but an incident doughnut mode with adapted beam waist to fit the ideal dipole-like radiant intensity as best as possible (dashed black curve),
d) conventional 4π setup (NA=0.95 in air) with linear polarization and constant intensity of the incident wave (red curves),
e) conventional 4π setup (NA=0.95 in air) with radial polarization and an incident intensity distribution producing a dipole-like radiant intensity between the two objectives (magenta curve),
f) conventional 4π setup (NA=0.95 in air) with radial polarization and an incident doughnut mode (for full and annular aperture, blue curves).

For other applications the width of the central maximum of the focus in lateral or axial direction is crucial. There, the parabolic mirror with dipole-like wave is quite good in the lateral direction but not as good as the conventional setups in axial direction. The smallest width in lateral direction is obtained by the conventional setup with radial polarization and an annular aperture. For the annular aperture one approaches the case of a conical wave such as the one produced by an axicon resulting in a Bessel beam [20]. The interference of two Bessel beams propagating in opposite directions can explain approximately the axial behaviour of the conventional 4π setup with annular aperture which has secondary maxima nearly as high as the central maximum which is the drawback of this system.

The axial sections (Fig. 9, right) show that the smallest central maximum along the optical axis is obtained with the conventional 4π setup and linear polarization, whereas the parabolic mirror has the broadest central maximum. However, the secondary maxima along the optical axis are very small for the parabolic mirror, whereas they are quite high for the conventional 4π setup, especially for the annular aperture and radial polarization.

In summary, the parabolic mirror produces a good confinement of the electric energy density at the focus in lateral as well as in axial direction. It also generates the highest possible electric field for a given incident light power. The conditions are illuminating with a plane wave front, radial polarization and a specially designed intensity distribution (see equation (25)) resulting in a dipole-like radiant intensity when approaching the focus. Consequently this setup should be especially useful for studying whether the absorption probability of a single atom can be nearly one for an incident photon with specially shaped state.

*3.4 Influence of aberrations*

It is well known that a widely opened parabolic mirror is very sensitive to misalignments of the incident wave. If the incident wave is only slightly off-axis, strong coma will be generated. This is shown in Fig. 10, where the cases of small off-axis angles of 0.05', 0.1' and 1' of the incident plane wave are compared with the ideal on-axis case. There, the electric energy density in the focal plane $|E|^2$ is given relative to the maximum value $|E_0|^2$ of the electric energy density of the incident wave. It can be seen that the maximum value of the relative electric energy density in the focus varies from $1.0 \cdot 10^{10}$ for the on-axis case to only $2.5 \cdot 10^8$ for the off-axis angle of one arc minute. For an off-axis angle of 0.1' the maximum value of the electric energy density is just half the value of the on-axis case. So, misalignments of the incident wave of order 0.1' are already too large to produce a sharp aberration-free focus.

For the conventional 4π setup with two aplanatic objectives the lenses themselves will not produce aberrations for small tilts of the incident wave because of the aplanatism. On the other hand, a tilt of the incident wave in only one of the lenses will produce a lateral shift of the focus and consequently there is no longer a common focus in the 4π setup. For a small tilt angle $\varphi$ of the incident wave and a focal length $f$ of the objective the lateral shift $\Delta x$ is:

$$\Delta x = f\varphi \qquad (28)$$

In our case with $f$=2.1 mm and $\varphi$=0.1' it is $\Delta x$=61 nm=0.24$\lambda$. Thus for the conventional 4π setup with aplanatic objectives a tilt of the incident wave of 0.1' will already disturb the interference pattern in the focal region as one of the foci is laterally shifted by nearly a quarter of the wavelength.

Figure 11 shows the electric energy density in the yz-plane of the focal region for different tilts in y-direction of one of the incident waves for the case of the conventional 4π setup illuminated with a plane wave with linear polarization in x-direction. By comparing the maxima of the different cases it can be seen that a decrease to about half the value of the on-axis case is obtained for a tilt angle of 0.2'. For the tilt angle of 1' the foci are nearly separated so that there is no longer interference between them and two normal foci of single objectives are obtained. In conclusion for the conventional 4π setup the requirements on the tilt of the incident waves are nearly the same as for the parabolic mirror even though the aplanatic objectives themselves do not introduce aberrations.

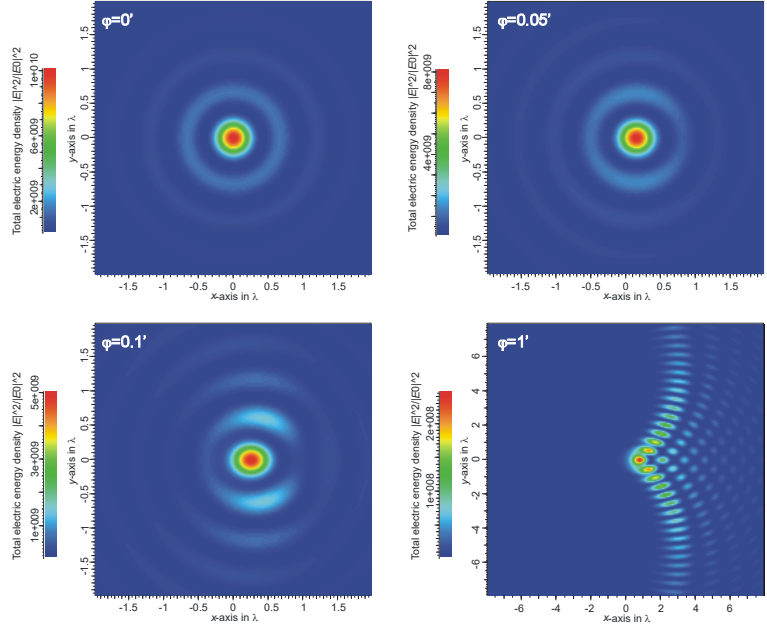

Fig. 10. Electric energy density in the focal plane of the parabolic mirror (mirror 1 using radial polarization and a dipole wave) for different off-axis angles $\varphi$ of the incident plane wave. The lateral scale is in multiples of the wavelength and the value $x=y=0$ is at the paraxial focus, respectively. Note the different lateral scales for $\varphi=1$' and the different scales of the electric energy density in all cases.

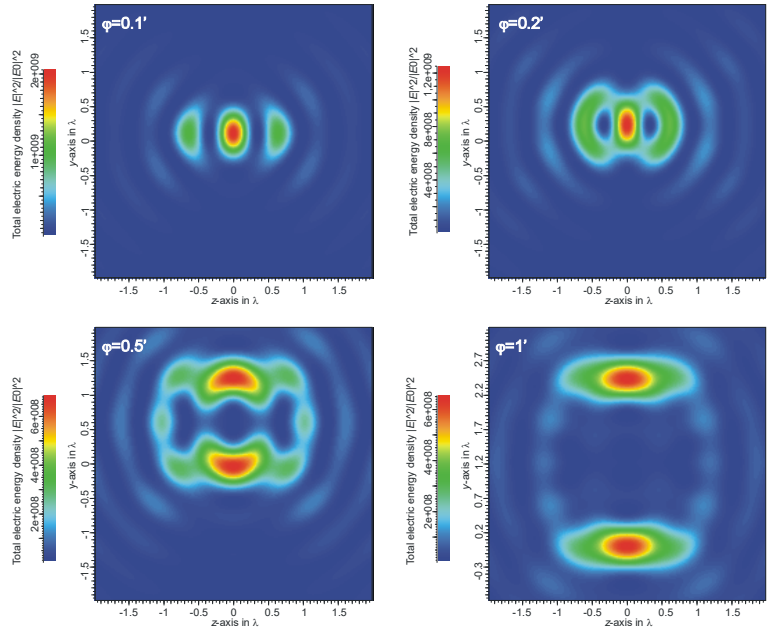

Fig. 11. Electric energy density in the yz-plane of the focal region of the conventional 4π setup (incident plane wave with linear polarization in x-direction) for different off-axis angles $\varphi$ in y-direction of one of the incident plane waves. The lateral scale is in multiples of the wavelength.

## 4. Summary

The paper first describes a numerical method for calculating the electric and magnetic energy density in the focal region of high numerical aperture optical systems (parabolic mirror and aplanatic lens) taking into account polarization. The method is based on the Debye integral [1,2]. In the case of a parabolic mirror it is also shown how misalignment induced aberrations can be included in the calculation.

The electric energy density in the focal region is calculated for a very deep parabolic mirror which is illuminated with a plane wave front, but specially shaped intensity distribution and radial polarization resulting in a dipole-like radiant intensity after reflection at the mirror [6]. It is shown that such an illumination results in a very small and well confined focus with the highest possible electric field relative to the incident light power. The very small secondary maxima along the optical axis may also be important for some applications.

The electric energy density in the focus of the parabolic mirror is also compared with that of a conventional $4\pi$ setup (NA=0.95 in air) with either full or annular aperture which is illuminated by a plane wave front with radial polarization and a doughnut shaped intensity distribution or by a homogeneous plane wave with linear polarization. All devices illuminated by radial polarization have only an axial component of the electric field in the focal plane, but the conventional $4\pi$ setup has higher secondary maxima along the optical axis, especially for the annular aperture. In the case of linear polarization the conventional $4\pi$ setup has only a lateral component (along the original direction of polarization) of the electric field in the focal plane.

However, since a parabolic mirror violates the sine condition there is strong coma if the incident wave is not exactly on-axis. For our parabolic mirror it is found that already a misalignment of 0.1 arc minutes off-axis angle is too large to allow for having a close to diffraction limited performance. For the conventional $4\pi$ setup the requirements on the tilt of the incident waves are nearly the same because a tilt of one of the incident waves will shift one of the foci of the two objectives laterally and so the interference between both foci will either be disturbed or there is no interference at all for large shifts.

The setup with the deep parabolic mirror producing a dipole-like radiant intensity can be used for investigating the coupling of a photon to an atom or ion with a linear dipole transition because the space inversed electric field distribution in the far field of a dipole wave is produced [19]. It should also be useful for confocal microscopy applications because of the very small secondary maxima along the optical axis which will result in a good contrast in axial direction.